\documentclass[12pt]{article}
\usepackage{amsmath,amssymb,amsthm,amsxtra,overpic,bbm,bm,epsfig}
\usepackage{color}
\usepackage{textcomp}
\usepackage[normalem]{ulem}
\usepackage{cite}
\usepackage{float}

\usepackage{hyperref}
\usepackage{url}
\usepackage{booktabs}

\allowdisplaybreaks[4]

\newcommand{\tabincell}[2]{\begin{tabular}{@{}#1@{}}#2\end{tabular}}
\begin{document}

\vspace{0.2cm}

\begin{center}
{\large\bf 
Terrestrial Matter Effects on Reactor Antineutrino Oscillations: Constant vs. Fluctuated Density Profiles
}
\end{center}

\vspace{0.2cm}

\begin{center}
{\bf Yu-Feng Li$^{1,5}$}
\footnote{liyufeng@ihep.ac.cn},
{\bf Andong Wang$^{2}$}
\footnote{adwang2013@ecut.edu.cn},
{\bf Ya Xu$^{3}$}
\footnote{xuya@mail.iggcas.ac.cn},
{\bf Jing-yu Zhu$^{4}$}
\footnote{zhujingyu@impcas.ac.cn (corresponding author)}
\\
{$^{1}$Institute of High Energy Physics,
Chinese Academy of Sciences, Beijing 100049, China \\
$^{2}$National Key Laboratory of Uranium Resources Exploration-Mining and Nuclear Remote Sensing, East China University of Technology, Nanchang 330013, China\\
$^{3}$Institute of Geology and Geophysics, Chinese academy of sciences, Beijing 100029, China \\
$^{4}$Institute of Modern Physics, Chinese Academy of Sciences, Lanzhou, 730000, China \\
$^{5}$School of Physical Sciences, University of Chinese Academy of Sciences, Beijing 100049, China }
\end{center}

\vspace{2cm}

\begin{abstract}

The JUNO Collaboration has recently released its first reactor antineutrino oscillation result, achieving unprecedented precision in the measurement of $\Delta m^2_{21}$ and $\sin^2\theta_{12}$. We emphasize that the accurate determination and modeling of the terrestrial matter density profile are fundamental for extracting the oscillation parameters and probing the neutrino mass ordering. 
This paper presents a realistic piecewise-constant model for the shallow crustal density profile along the baselines from Taishan and Yangjiang to the experimental hall, based on geological and petrophysical information. The uncertainty in the density profiles arises from variations in the density and length of each segment, both of which are conservatively estimated to be 10\%.
A careful comparison of constant and fluctuated density profiles is provided and the implications for the precision measurement of oscillation parameters are discussed. Finally, we also discuss the prospect of shallow crust tomography in future reactor neutrino experiments.
\end{abstract}

\newpage

\section{Introduction}
The discovery of the neutrino oscillation phenomena~\cite{Kajita:2016cak,McDonald:2016ixn}, which demonstrated that neutrinos have masses and that lepton flavors are mixed, marks a significant milestone in particle physics. Over the past two decades, numerous solar, atmospheric, reactor, and accelerator neutrino experiments have collected a substantial number of neutrino events~\cite{ParticleDataGroup:2024cfk} and established a coherent picture of the three-flavor neutrino oscillation framework~\cite{Capozzi:2025wyn,Esteban:2024eli,deSalas:2020pgw}, which is described by the
$3\times 3$ Pontecorvo--Maki--Nakagawa--Sakata (PMNS) ~\cite{Pontecorvo:1957cp,Maki:1962mu,Pontecorvo:1967fh} matrix:
\begin{eqnarray}
U = \left( \begin{matrix}
 c_{12} c_{13} & s_{12} c_{13} & s_{13}e^{-{\rm i } \delta} \cr
 -s_{12} c_{23} - c_{12} s_{13} s_{23}e^{{\rm i } \delta} & 
 c_{12} c_{23} - s_{12} s_{13} s_{23}e^{{\rm i } \delta} & 
 c_{13} s_{23} \cr
 s_{12} s_{23} - c_{12} s_{13} c_{23} e^{{\rm i } \delta} & 
 -c_{12} s_{23} - s_{12} s_{13} c_{23}e^{{\rm i } \delta} & 
 c_{13} c_{23}
\cr \end{matrix} \right) ,
\label{eq:PMNS}
\end{eqnarray}
with $c_{ij} \equiv \cos\theta_{ij}$ and $s_{ij} \equiv \sin\theta_{ij}$ ($ij = 12,13,23$), and $\delta$ the Dirac CP-violating phase, and three mass-squared differences:
\begin{eqnarray}
\Delta m^2_{ij} = m_i^2 - m_j^2\;,
\label{eq:masss}
\end{eqnarray}
where $(i,j = 1,2,3,\, i>j)$ and $m_i$ is the mass of the $i$-th mass eigenstate $\nu_i$.

The three-flavor oscillation framework has been successfully measured and tested to within a few percent~\cite{Capozzi:2025wyn,Esteban:2024eli,deSalas:2020pgw}. However, several key scientific questions remain unanswered. Most notably, the ordering of the neutrino mass spectrum—normal mass ordering (NMO: $m_1<m_2<m_3$) versus inverted mass ordering (IMO: $m_3<m_1<m_2$)—is still unknown, as is the value of the Dirac CP-violating phase $\delta$.
In addition, the absolute neutrino mass scale and the nature of neutrino masses—specifically, whether neutrinos are Dirac or Majorana particles—remain unknown. They can be probed by $\beta$ decay and neutrinoless double-$\beta$-decay experiments, as well as cosmological observations \cite{KATRIN:2021uub, Agostini:2022zub, KamLAND-Zen:2024eml, Planck:2018vyg}. 

The global neutrino program in the coming decades is organized around these promising frontiers. Next-generation long-baseline accelerator experiments such as Hyper-Kamiokande~\cite{Hyper-Kamiokande:2018} and DUNE~\cite{DUNE:2020} aim to precisely measure the atmospheric oscillation parameters and the CP-violating phase $\delta$. On the reactor side, the Jiangmen Underground Neutrino Observatory (JUNO)~\cite{JUNO:2015zny,JUNO:2022} is a next-generation oscillation experiment aiming at determining the neutrino mass ordering~\cite{Li:2013zyd,JUNO:2024jaw} and measuring oscillation parameters with sub-percent precision~\cite{JUNO:2022mxj}. 
All of these next-generation neutrino oscillation experiments will operate in an era of high precision and require better control of systematic uncertainties on both the theoretical and experimental sides.

The JUNO experiment, the first of these next-generation flagship experiments to begin operations, has published its initial reactor antineutrino oscillation results~\cite{JUNO:2025gmd,JUNO:2025fpc}. It achieved unprecedented precision in measuring $\Delta m^2_{21}$ and $\sin^2\theta_{12}$, with uncertainties of 1.55\% and 2.81\%, respectively. With this accuracy level, the  
corrections due to terrestrial matter effects~\cite{Wolfenstein:1977ue,Mikheev:1986wj} on $\Delta m^2_{21}$ and $\sin^2\theta_{12}$ will be at the percent level~\cite{Li:2016txk,Capozzi:2013psa},
and cannot be neglected in the analysis.
In previous analyses, a constant density profile, characterized by an average crustal density, was assumed and deemed reasonably accurate. However, since the majority of the neutrino propagation trajectories are relatively shallow, the density variation may be significant depending on the local geological conditions. Therefore, it is important to evaluate the effect of a variable density profile on the survival probability and its potential impact on the oscillation parameters.

In this work, we focus on the impact of realistic crustal density variations along the JUNO baselines. By utilizing geological maps and petrophysical data from South China, we construct a piecewise-constant matter density model along the baselines extending from the Taishan and Yangjiang nuclear power plants to the experimental hall. We evaluate the uncertainties in the density profiles by incorporating detailed local information regarding crustal density and geometric baseline variations in each segment.
Using an efficient semi-analytical method based on the Cayley-Hamilton theorem, we compare the differences in neutrino survival probabilities between the variable density profile and its constant density approximation, and discuss the potential impact on future measurements of oscillation parameters. Finally, we explore the possibility of shallow Earth crust tomography in future reactor neutrino experiments.

The remainder of this paper is organized as follows: In Sec.~2, we construct the density profile model along the Taishan--JUNO and Yangjiang--JUNO baselines using geological information. In Sec.~3, we provide a brief recap of three-flavor neutrino oscillations in matter and introduce the semi-analytical formalism used to analyze piecewise-constant density profiles. Sec.~4 presents our numerical results for the survival probabilities and illustrates the impacts of different choices of density profiles. In Sec.~5, we further investigate the impact of possible anomalous density perturbations, such as localized water- or air-like structures along the baselines, and discuss their implications for shallow crust tomography with reactor antineutrinos. Finally, our conclusions and a brief outlook are provided in Sec.~6.

\section{Matter density profile along the baselines}
\label{sec:matter-profile}

\subsection{Geological model and baseline discretizations}
\label{subsec:geo-model}

The JUNO detector is located about $750~\mathrm{m}$ underground, whereas the
reactors are slightly above sea level. Reactor antineutrinos thus propagate
predominantly through the upper crust, over relatively shallow depths. In this
regime, lateral variations of the density are dominated by changes in outcropping
strata and lithology rather than by large-scale crust–mantle structures. Under
these conditions, it is reasonable to approximate each surface geological unit by
a representative bulk density; residual small-scale variations are expected at the
level of a few percent. In the following sections we adopt segment-wise fluctuations
up to $\pm 10\%$ as a deliberately conservative bound that comfortably covers these
geophysical uncertainties.

\begin{figure}[t]
  \centering
  \begin{minipage}[c]{0.5\textwidth}
    \centering
    \includegraphics[width=\textwidth]{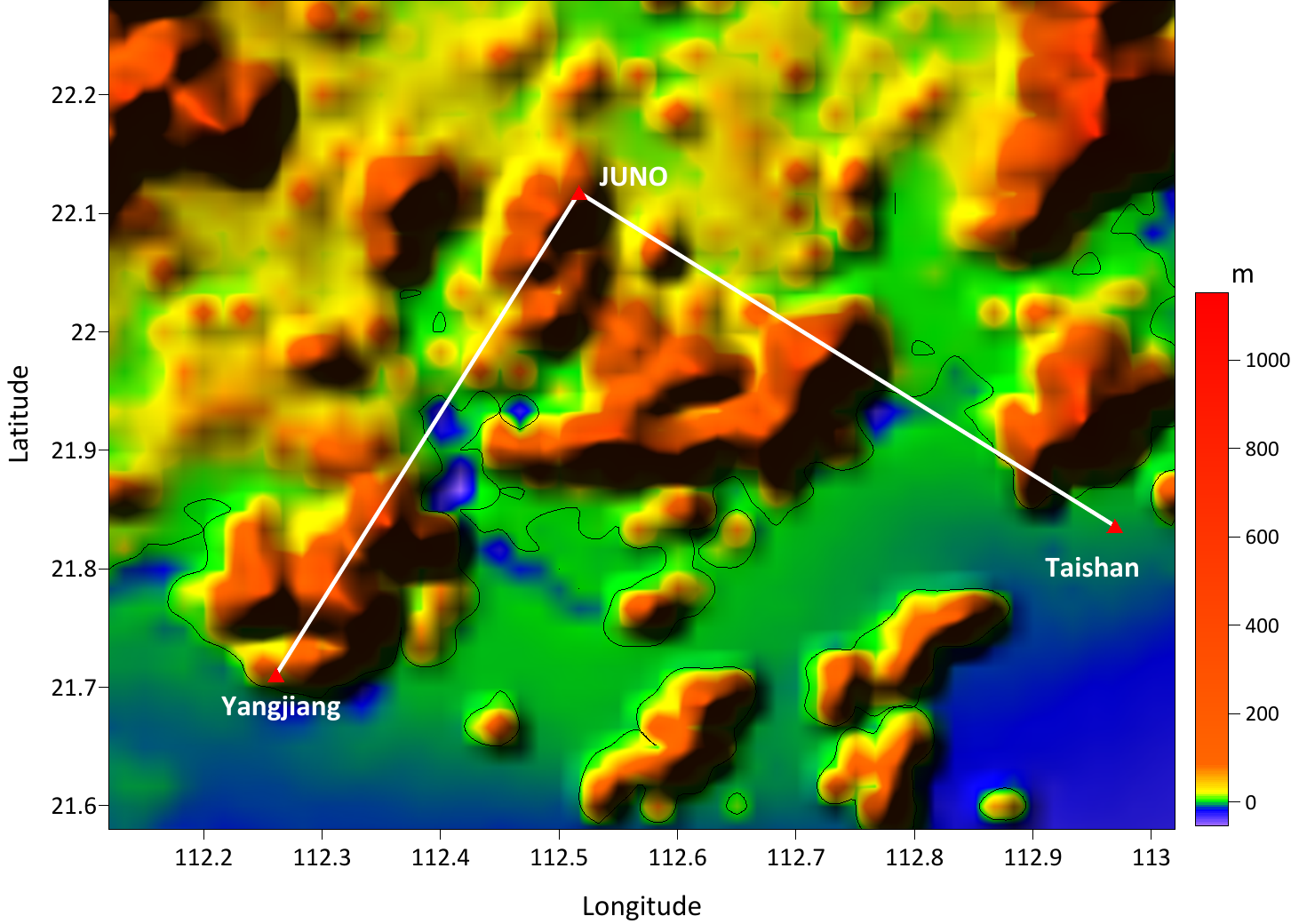} 
  \end{minipage}
  \hspace{0.02\textwidth}
  \begin{minipage}[c]{0.38\textwidth}
    \centering
    \includegraphics[width=\textwidth]{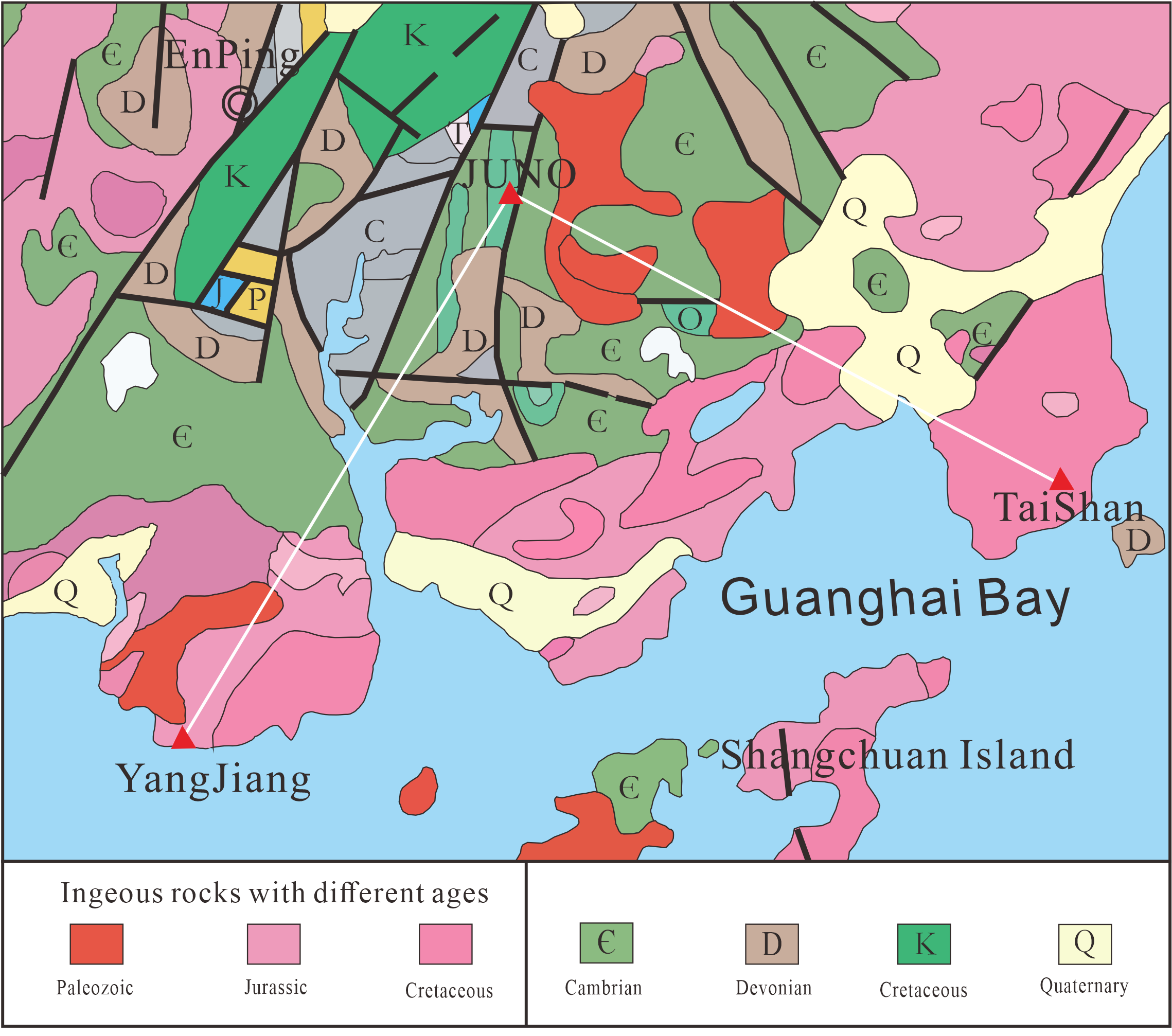}  
  \end{minipage}
  \caption{Topographic (left) and geological (right) maps of the terrain and crustal structure traversed by the reactor antineutrino oscillation baselines from the Taishan (TS) and Yangjiang (YJ) nuclear power plants to the JUNO detector. The white lines indicate the TS--JUNO and YJ--JUNO baselines.
}
  \label{fig:topo_geo_JUNO}
\end{figure}

We briefly summarize the construction of our density profile model. Figure~\ref{fig:topo_geo_JUNO} shows the topographic and geological maps of the JUNO detector site at Jiangmen and the two reactor complexes,
Taishan (TS) and Yangjiang (YJ). The topographic map illustrates the surface elevation along the two
baselines, while the geological map displays the distribution of surface rock units in South China. The
white lines indicate the straight baselines from the reactor cores to JUNO, which traverse a
variety of surface geological units. The density structure along each baseline is inferred from these mapped
geological units together with the petrophysical properties of the corresponding strata and rock types
\cite{deng2019rock}.

To model the matter density along these baselines, we extract the strata and
lithology units intersected by the straight line connecting each reactor complex
to the JUNO detector from the geological map. Wherever the baseline crosses a
boundary between two mapped surface geological units, we define a new segment. In
this way, the path from Taishan to JUNO is divided into 12 rock segments, and the
path from Yangjiang to JUNO into 14 rock segments. The length of each segment is
determined by the distance between consecutive intersections with geological
boundaries.

It is worth noting that the actual oscillation baselines are slightly tilted with
respect to the local ground surface, since the reactor cores are located about
$2~\mathrm{m}$ above ground level, while the JUNO detector is situated at a depth of
about $750~\mathrm{m}$ underground. Along the straight line connecting a reactor core to
the JUNO detector, the antineutrinos therefore first propagate over a short
distance in air before entering the shallow crust. In our piecewise-constant
description we account for this by adding an initial segment with density
$\rho = 0$ in front of the rock segments listed in Tables~1 and 2. As a result,
although the geological sampling along the TS--JUNO (YJ--JUNO) baseline yields
12 (14) rock segments, the effective profiles used in the oscillation
calculations consist of $N_{\rm TS}=13$ and $N_{\rm YJ}=15$ segments in total,
including the initial air segment. This segmented representation of the baselines
will be used in the following both for the discussion of matter effects and for
comparison with previous studies based on a constant-density approximation.

Previous studies have conducted extensive tests and summarized the density
characteristics of typical strata and rocks in South China, as compiled in
Ref.~\cite{deng2019rock}. The densities assigned to each segment are based on typical
rock densities for the corresponding lithology along the JUNO baselines. The
resulting segment lengths, densities and rock types for TS--JUNO and YJ--JUNO
are summarized in Tables~1 and 2, respectively. The surface elevation and the
associated surface density along each baseline are plotted in Figs.~2 and~3.
These figures illustrate that the density variations are relatively modest,
typically at the level of a few percent around a mean crustal density of
$\sim 2.5~\mathrm{g/cm^3}$.

\subsection{Fluctuations of the matter density profile}
\label{subsec:profile-fluct}

In the following we discuss how to model realistic fluctuations of the matter
density profile along the Taishan--JUNO and Yangjiang--JUNO baselines, and
clarify their impact on the effective average density that enters the
neutrino oscillation probabilities.

For definiteness, we denote the reference piecewise-constant profile obtained
from the geological information in Sec.~\ref{subsec:geo-model} by
$\{\rho_i^{(0)}, L_i^{(0)}\}$, where $\rho_i^{(0)}$ and $L_i^{(0)}$ are
the density and length of the $i$-th segment, and
$\sum_i L_i^{(0)} = L$ is the total baseline length.
As discussed in Sec.~\ref{subsec:geo-model}, the Taishan--JUNO baseline is
divided into $N_{\rm TS}=13$ segments and the Yangjiang--JUNO baseline into
$N_{\rm YJ}=15$ segments. The first segment in both cases corresponds to
propagation in air, and thus its matter density is set to zero and not
allowed to fluctuate.

\begin{table}[H]
\centering
\begin{tabular}{p{2.2cm}p{1.5cm}p{10cm}}
\toprule
\tabincell{l}{Distance \\(km)} & \tabincell{l}{Density\\ (${\rm g/cm^3}$)} & \tabincell{l}{Rock types}\\
\midrule
11.5 & 2.61 & Cretaceous: biotite granite \\
14.8 & 2.49 & Quaternary \\
20.6 & 2.49 & Quaternary \\
25.4 & 2.61 & Cretaceous: biotite granite\\
27.5 & 2.52 & Cambrian: interbedded sandstone--shale or slate, with limestone, sandy gravel, and stone coal\\
32.6 & 2.54 & Paleozoic: plagioclase-rich granite\\
39.2 & 2.52 & Cambrian: interbedded sandstone--shale or slate, with limestone, sandy gravel, and stone coal\\
41.7 & 2.55 & Paleozoic: rhyolite porphyry\\
46.7 & 2.53 & Cambrian: interbedded sandstone--shale or slate, with limestone, sandy gravel, and stone coal \\
48.5 & 2.62 & Paleozoic: granodiorite\\
52.0 & 2.52 & Cambrian: interbedded sandstone--shale or slate, with limestone, sandy gravel, and stone coal\\
52.7 & 2.51 & Ordovician: graptolite shale, siliceous shale, sandstone, with interbedded conglomerate\\
\bottomrule
\end{tabular}
\caption{\label{tab:segments-TS}
Matter density sampling and corresponding rock types along the baseline from Taishan (TS) to JUNO. The distances are rounded to one decimal place.}
\end{table}

\begin{table}[H]
\centering
\begin{tabular}{p{2.2cm}p{1.5cm}p{10cm}}
\toprule
\tabincell{l}{Distance \\(km)} & \tabincell{l}{Density\\ (${\rm g/cm^3}$)} & \tabincell{l}{Rock types}\\
\midrule
11.5 & 2.60 & Jurassic: biotite granite, monzogranite\\
15.0 & 2.61 & Paleozoic: granite, gneissic granite\\
16.1 & 2.55 & Jurassic: biotite granite, monzogranite \\
18.9 & 2.58 & Jurassic: granite porphyry\\
19.8 & 2.61 & Cretaceous: quartz syenite \\
26.5 & 2.52 & Cambrian: interbedded sandstone--shale or slate, with limestone, sandy gravel, and stone coal \\
29.5 & 2.57 & Unmapped area, inferred to be Cambrian or Middle to Upper Devonian\\
33.3 & 2.52 & Cambrian: interbedded sandstone--shale or slate, with limestone, sandy gravel, and stone coal\\
35.2 & 2.57 & Devonian\\
40.7 & 2.52 & Cambrian: interbedded sandstone--shale or slate, with limestone, sandy gravel, and stone coal \\
43.7 & 2.51 & Ordovician: graptolite shale, siliceous shale, sandstone, with interbedded conglomerate\\
47.4 & 2.57 & Devonian \\
48.9 & 2.52 & Cambrian: interbedded sandstone--shale or slate, with limestone, sandy gravel, and stone coal\\
52.5 & 2.52 & Ordovician: graptolite shale, siliceous shale, sandstone, with interbedded conglomerate\\
\bottomrule
\end{tabular}
\caption{\label{tab:segments-YJ}
Matter density sampling and corresponding rock types along the baseline from Yangjiang (YJ) to JUNO. The distances are rounded to one decimal place.}
\end{table}

The baseline-averaged matter density corresponding to the reference profile
is defined as
\begin{equation}
  \label{eq:rho-average-def}
  \rho^{\rm Average}
  \;=\;
  \frac{1}{L}\sum_{i=1}^{N} \rho_i^{(0)}\,L_i^{(0)} \;,
\end{equation}
with $N=N_{\rm TS}=13$ or $N=N_{\rm YJ}=15$ for the two baselines.
Using the segment densities and lengths in Tables~1 and 2 we obtain
$\rho_{\rm TS}^{\rm Average} \simeq 2.541~{\rm g/cm^3}$ and
$\rho_{\rm YJ}^{\rm Average} \simeq 2.554~{\rm g/cm^3}$,
which will be used as the benchmark input of the constant-density approximation
in Sec.~4.

In the numerical calculations of Sec.~4, this piecewise-constant profile is
used as the reference density model. To account for possible uncertainties in
the geological interpretation and for local heterogeneities that are not
resolved by the surface map, we adopt conservative fluctuation ranges in both
density and segment length. Specifically, we allow the density in each segment
to vary independently within $\pm 10\%$ of its reference value, and the
effective lengths of the segments to fluctuate within $\pm 10\%$ while
keeping the total baseline length fixed, as detailed below.

\begin{figure}[H]
    \centering
    \includegraphics[width=0.9\linewidth]{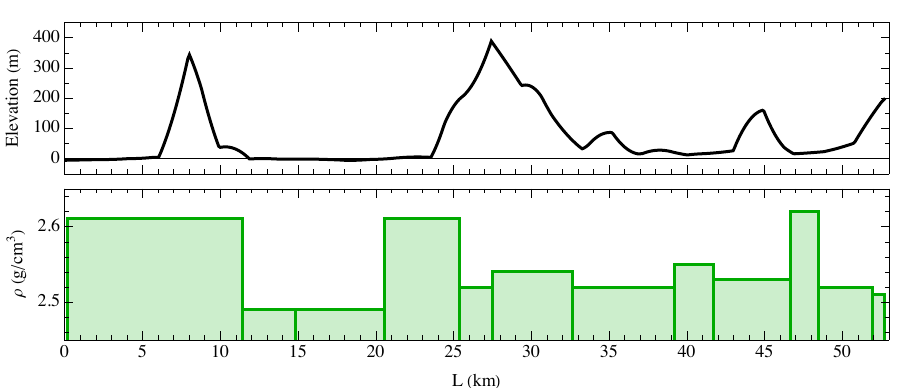}
    \caption{Geological sampling along the TS–JUNO baseline. The upper panel shows the surface elevation as a function of distance, and the lower panel shows the corresponding matter density of surface rocks.
}
    \label{fig:TStoJUNO}
\end{figure}

\begin{figure}[H]
    \centering
    \includegraphics[width=0.9\linewidth]{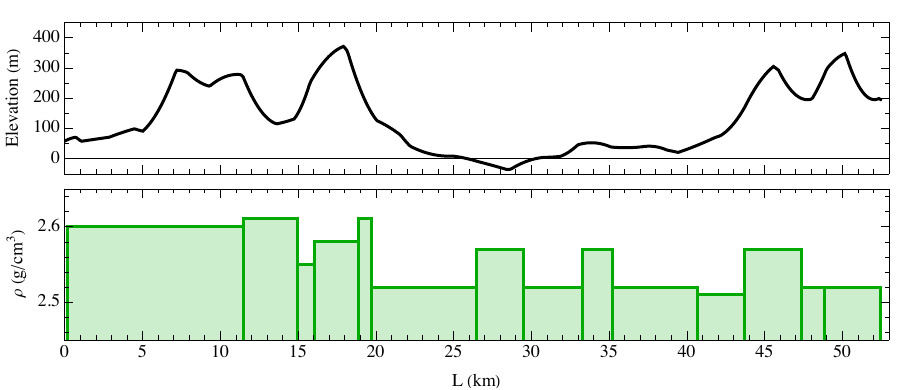}
    \caption{Geological sampling along the YJ–JUNO baseline. The upper panel shows the surface elevation as a function of distance, and the lower panel shows the corresponding matter density of surface rocks.}
    \label{fig:YJtoJUNO}
\end{figure}

\vspace{2mm}
\noindent\textbf{(i) Fluctuations of segment densities.}
To model the uncertainty of the crustal density itself, we allow each segment
density to fluctuate around its reference value,
\begin{equation}
  \label{eq:rho-fluct}
  \rho_i
  \;=\;
  \rho_i^{(0)}\,(1+\delta\rho_i) \;,
  \qquad
  |\delta\rho_i|\leq 10\% \;,
\end{equation}
where $\delta\rho_i$ are treated as independent random variables uniformly
distributed in the interval $[-0.1,\,0.1]$.\footnote{%
  In the numerical analysis of Sec.~4 we include the density fluctuations
  only for the rock segments. The density of the air segment is kept fixed at
  $\rho=0$, although its geometric length is allowed to fluctuate together
  with the other segments, see below.}
The total baseline length is kept fixed, $L_i=L_i^{(0)}$, in this case.
For a given realization of $\{\delta\rho_i\}$ the corresponding average
density reads
\begin{equation}
  \label{eq:rho-avg-rho-only}
  \rho^{\rm Average}_\rho
  \;=\;
  \frac{1}{L}\sum_i \rho_i L_i^{(0)}
  \;=\;
  \rho^{\rm Average}
  + \frac{1}{L}\sum_i \rho_i^{(0)} L_i^{(0)} \delta\rho_i \;,
\end{equation}
so that the relative deviation from the reference value,
$\Delta\rho^{\rm Average}_\rho \equiv \rho^{\rm Average}_\rho -
\rho^{\rm Average}$, is directly given by a weighted average of the segment
fluctuations $\delta\rho_i$.
In the numerical scan in Sec.~4 we find that for
$|\delta\rho_i|\leq 10\%$ the typical (one-standard-deviation) fluctuation
of $\rho^{\rm Average}$ is of order $1.5\%$ for both baselines, and
$95\%$ of the random realizations lie within about $\pm 3\%$
around the reference value.  Thus, even if each segment is allowed to vary
independently within $\pm 10\%$, the baseline-averaged density remains
remarkably stable at the percent level.

It is also instructive to consider the opposite limiting case in which the
density fluctuations of all segments are fully correlated. In this case we may
write
\begin{equation}
  \delta\rho_i \equiv \delta \,,
  \qquad |\delta| \leq 10\% \,,
\end{equation}
so that all segment densities are rescaled by the same factor,
$\rho_i = \rho_i^{(0)} (1+\delta)$. The corresponding average density is then
given by
\begin{equation}
  \rho^{\rm Average}_{\rm corr}
  \;=\;
  \frac{1}{L}\sum_i \rho_i^{(0)} (1+\delta)\,L_i^{(0)}
  \;=\;
  (1+\delta)\,\rho^{\rm Average} \,,
\end{equation}
and the relative deviation of the average density is simply
\begin{equation}
  \frac{\rho^{\rm Average}_{\rm corr} - \rho^{\rm Average}}
       {\rho^{\rm Average}}
  \;=\; \delta \,.
\end{equation}
If the common fluctuation $\delta$ is assumed to be uniformly distributed in
the interval $[-0.1,\,0.1]$, the associated one-standard-deviation fluctuation
is $\sigma_\delta = 0.1/\sqrt{3} \simeq 5.8\%$, and the full allowed range of
$\rho^{\rm Average}$ extends up to $\pm 10\%$ around the reference value. This
fully correlated scenario corresponds to a global mis-normalization of the
crustal density scale, rather than to local heterogeneities between different
segments. In the following we treat it as a conservative upper bound on possible
shifts of the effective average density; its impact on the oscillation
probabilities is effectively illustrated by the constant-density variations
considered in Sec.~4.

\vspace{2mm}
\noindent\textbf{(ii) Fluctuations of segment lengths.}
The uncertainties of the segment boundaries correspond to uncertainties in the
segment lengths. A physically motivated way to describe them is to regard
the internal interfaces along the baseline as being known only up to a
finite longitudinal resolution.  Denoting the original positions of the
interfaces by $x_k^{(0)}$ ($k=1,\dots,N+1$), with
$x_1^{(0)}=0$, $x_{N+1}^{(0)}=L$ and
$L_i^{(0)} = x_{i+1}^{(0)} - x_i^{(0)}$, we allow each internal point
($k=2,\dots,N$) to move independently within a small interval around its
reference position,
\begin{equation}
  \label{eq:xk-fluct}
  x_k \;=\; x_k^{(0)} + \Delta x_k \;,
  \qquad
  -\,\alpha\,L_{k-1}^{(0)} \;\leq\; \Delta x_k \;\leq\; \alpha\,L_k^{(0)} \;,
\end{equation}
where we take $\alpha=0.1$ as a conservative estimate.  The new segment
lengths are then given by
\begin{equation}
  \label{eq:Li-fluct-from-xk}
  L_i \;=\; x_{i+1} - x_i
  \;=\; L_i^{(0)} + \delta L_i \;,
  \qquad
  \delta L_i \equiv \Delta x_{i+1}-\Delta x_i \;.
\end{equation}
By construction the baseline length remains exactly fixed,
$\sum_i L_i = x_{N+1}-x_1 = L$, and neighbouring segments are
anti-correlated: if an interface moves to the right, the segment on its
left becomes shorter and that on its right becomes longer.  We stress that
this picture corresponds closely to the physical origin of the geometric
uncertainty, namely the limited knowledge of where exactly the geological
interfaces intersect the oscillation baseline.

In this case we keep the segment densities fixed at their reference values,
$\rho_i = \rho_i^{(0)}$, and the average density becomes
\begin{equation}
  \label{eq:rho-avg-L-only}
  \rho^{\rm Average}_L
  \;=\;
  \frac{1}{L}\sum_i \rho_i^{(0)} L_i
  \;=\;
  \rho^{\rm Average}
  + \frac{1}{L}\sum_i \big(\rho_i^{(0)}-\rho^{\rm Average}\big)\,\delta L_i \;.
\end{equation}
Equation~\eqref{eq:rho-avg-L-only} makes it clear why the impact of the
segment-length fluctuations on $\rho^{\rm Average}$ is much smaller than
that of the density fluctuations in Eq.~\eqref{eq:rho-avg-rho-only}.
Only the deviations of each segment density from the already narrow band
around $\rho^{\rm Average}$ enter, and the contributions from the different
segments tend to cancel because the $\delta L_i$ are constrained by
$\sum_i \delta L_i = 0$ and are strongly anti-correlated through
Eq.~\eqref{eq:Li-fluct-from-xk}.  Numerically we find that for
$|\delta L_i|/L_i^{(0)}\lesssim 10\%$ the typical fluctuation of
$\rho^{\rm Average}$ induced by the length variations alone is only of
order $10^{-3}$, i.e.\ at the level of a few $0.01\%$, for both baselines.
Therefore the geometrical uncertainty of the segment lengths has a
negligible impact on the effective average matter density.\footnote{%
  As a cross-check we have also considered a simpler scheme in which each
  segment length is independently rescaled according to
  $L_i \to L_i^{(0)}(1+\delta L_i)$ with $|\delta L_i|\leq10\%$, followed
  by a global rescaling of all $L_i$ such that the total baseline length $L$
  is restored.  This alternative model slightly overestimates the effect of
  length fluctuations, but still yields fluctuations of
  $\rho^{\rm Average}$ at the level of $\sim 0.05\%$, fully consistent with
  the more realistic interface-displacement model described by
  Eqs.~\eqref{eq:xk-fluct} and \eqref{eq:Li-fluct-from-xk}.}

\vspace{2mm}
\noindent\textbf{(iii) Combined fluctuations of densities and lengths.}
Finally, in order to mimic a conservative and yet realistic matter profile
uncertainty, we allow both the segment densities and the segment lengths to
fluctuate simultaneously according to Eqs.~\eqref{eq:rho-fluct},
\eqref{eq:xk-fluct} and \eqref{eq:Li-fluct-from-xk}.  To leading order in
the small quantities $\delta\rho_i$ and $\delta L_i$, the deviation of the
average density can be written as
\begin{equation}
  \label{eq:rho-avg-both}
  \rho^{\rm Average}_{\rho+L}
  \;\simeq\;
  \rho^{\rm Average}_\rho + \big(\rho^{\rm Average}_L - \rho^{\rm Average}\big)
  \;,
\end{equation}
so that the variances of the two contributions approximately add.  Since
the length-induced part is suppressed by the small spread of the segment
densities around $\rho^{\rm Average}$ and by cancellations between
neighbouring segments, its variance is much smaller than that induced by
the density fluctuations.  Therefore the distribution of
$\rho^{\rm Average}_{\rho+L}$ is practically indistinguishable from that
obtained in the density-only case in Eq.~\eqref{eq:rho-avg-rho-only}.
In particular, the typical size of the fluctuation of the average density
for $|\delta\rho_i|\leq 10\%$ remains at the level of $\mathcal{O}(1\%)$,
while the additional uncertainty from the segment-length fluctuations stays
at the $\mathcal{O}(10^{-3})$ level.  As a consequence, in the discussion
of matter effects on the reactor antineutrino oscillation probabilities in
Sec.~4 we will focus on the impact of the density fluctuations, and treat
the uncertainty due to the geometric segmentation as a subdominant effect.

\section{Neutrino oscillations in matter}
\setcounter{equation}{0}

\noindent

In this section, we recap the fundamental formulas for three-flavor neutrino oscillations in matter and introduce an efficient semi-analytical method for exact calculations.

In vacuum, the neutrino oscillations are governed by a Schr\"odinger-like evolution equation in the flavor basis,
\begin{equation}
 {\rm i} \frac{{\rm d}}{{\rm d}x} \nu_\alpha(x) \;=\; H_{\rm vac}\,\nu_\alpha(x)\,,
 \label{eq:eom-vac}
\end{equation}
where $\nu_\alpha = (\nu_e,\nu_\mu,\nu_\tau)^T$ denotes the flavor eigenstate vector and $x$ the distance neutrinos have travelled. The vacuum Hamiltonian reads
\begin{equation}
 H_{\rm vac} \;=\; \frac{1}{2E_\nu} \left[ U
 \begin{pmatrix}
 0 & 0 & 0 \\
 0 & \Delta m_{21}^2 & 0 \\
 0 & 0 & \Delta m_{31}^2
 \end{pmatrix}
 U^\dagger 
 \right] ,
 \label{eq:Hamiltonian-vac}
\end{equation}
where $E_\nu$ is the neutrino energy.

For reactor antineutrino experiments, we are interested in the survival probability of electron antineutrinos, $P(\bar\nu_e\to\bar\nu_e)$, which can be written in a compact form as~\cite{Xing:2018zno,JUNO:2022mxj}:
\begin{equation}
\begin{aligned}
P(\bar{\nu}_{e} \rightarrow \bar{\nu}_{e}) = 1
& - \sin^{2} 2\theta_{12} \cos^{4} \theta_{13} \sin^{2} \frac{\Delta m_{21}^{2} L}{4E_{\nu}} \\
& - \frac{1}{2} \sin^{2} 2\theta_{13} \left( \sin^{2} \frac{\Delta m_{31}^{2} L}{4 E_{\nu}} + \sin^{2} \frac{\Delta m_{32}^{2} L}{4 E_{\nu}} \right) \\
& - \frac{1}{2} \cos 2\theta_{12} \sin^{2} 2\theta_{13} \sin \frac{\Delta m_{21}^{2} L}{4 E_{\nu}} \sin \frac{(\Delta m_{31}^{2} + \Delta m_{32}^{2}) L}{4 E_{\nu}}\,,
\end{aligned}
\label{eq:Pee-vac}
\end{equation}
where $L$ denotes the baseline length. 

When neutrinos propagate in matter, coherent forward scattering on electrons induces an effective potential~\cite{Wolfenstein:1977ue,Mikheev:1986wj} that modifies the Hamiltonian. Therefore, the effective Hamiltonian of antineutrinos in the flavor basis reads
\begin{equation}
 H(x) \;=\; \frac{1}{2E_\nu} \left[ U
 \begin{pmatrix}
 0 & 0 & 0 \\
 0 & \Delta m_{21}^2 & 0 \\
 0 & 0 & \Delta m_{31}^2
 \end{pmatrix}
 U^\dagger \,-\,
 \begin{pmatrix}
 A(x) & 0 & 0 \\
 0 & 0 & 0 \\
 0 & 0 & 0
 \end{pmatrix}
 \right] ,
 \label{eq:Hamiltonian-matter}
\end{equation}
with
\begin{equation}
 A(x) \;=\; 2\sqrt{2}\,G_{\rm F}\,N_e(x)\,E_\nu \simeq 1.52 \times
 10^{-7}~{\rm eV^2} \times Y_e \left[\frac{\rho (x)}{\rm g\cdot cm^{-3}}\right] \times \left[\frac{E_{\nu}}{\rm MeV}\right]\,.
\end{equation}
Here $G_{\rm F}$ is the Fermi constant, $N_e(x)$ the electron number density, $\rho(x)$ the matter density, and $Y_e$ the electron fraction per nucleon, which is typically $Y_e\simeq 0.5$ in the crust of the Earth.

Under the approximation of a constant matter density along the baseline, we have 
$\rho(x) \simeq \bar\rho$ and $A(x) \simeq \bar A = 2\sqrt{2}\,G_{\rm F}\,\bar N_e\,E_\nu$, which leads to an $x$-independent Hamiltonian $H^{\rm C}$ of Eq.~\ref{eq:Hamiltonian-matter}. The evolution operator over a baseline $L$ then takes the simple form
$ S^{\rm C}(L) \;=\; \exp\!\left[-\,{\rm i}\,H^{\rm C}\,L \right]$, and the corresponding $\bar\nu_e$ survival probability can be calculated as 
$ P_{ee}^{\rm C}(E_\nu,L) \;=\; \big|S^{\rm C}_{ee}(L)\big|^2 $.

Approximate analytic expressions for $P_{ee}^{\rm C}$ in the three-flavor neutrino oscillation framework, valid for reactor experiments, can be found in Ref.~\cite{Li:2016txk}, where the leading matter corrections were shown to be relevant for precision measurements of $\Delta m^2_{21}$ and $\sin^2\theta_{12}$ and mass ordering. The exact three-flavor oscillation probabilities in constant-density matter for arbitrary flavor transitions can be written in fully analytical form, see for example Refs.~\cite{Barger:1980tf,Zaglauer:1988gz,Naumov:1991ju,Toshev:1991ku,Xing:2000gg,Ohlsson:1999xb,Freund:2001pn,Zhang:2004hf,Xing:2018lob}, while a variety of compact approximate formulas tailored to different experimental regimes are available in Refs.~\cite{Akhmedov:2004ny,Li:2015oal,Li:2016txk,Xing:2016ymg,Luo:2019efb,Wang:2019dal,Wang:2019yfp,Xing:2019owb,Zhu:2020wuy}.


In a realistic and accurate manner, the crust consists of different geological units with different densities, and the baselines from each reactor complex to the detector cross multiple rock formations. To take this into account, we consider a position-dependent matter density profile $\rho(x)$ obtained from local geological surveys and geophysical data. For practical calculations, it is convenient to approximate $\rho(x)$ by a set of $N$ segments with piecewise constant densities,
\begin{equation}
 \rho(x) \;\simeq\; \rho_i\,, \qquad x_{i-1} < x < x_{i}\,, \qquad i = 1,\dots,N\,,
 \label{eq:segment-density}
\end{equation}
with segment lengths $L_i = x_i - x_{i-1}$ satisfying $\sum_{i=1}^N L_i = L$.

In each segment the Hamiltonian $H_i$ is constant, so the evolution operator in the corresponding segment is
\begin{equation}
 S_i \;=\; \exp\!\left[-\,{\rm i}\,H_i\,L_i\right] = \exp \left[-{\rm i} M_i L_i/(2E_{\nu})\right],
\end{equation}
where $M_i$ is the effective mass-squared matrix in matter. According to the Cayley--Hamilton theorem, the matrix exponential $S_i$ for a $3\times 3$ matrix can be expressed as a quadratic polynomial in $M_i$~\cite{Moler1978,Ohlsson:1999xb,Li:2016pzm},
\begin{equation}
S_i = s_0 I + s_1 M_i + s_2 M_i^2\;,
\end{equation}
where $I$ is the $3\times 3$ identity matrix, and the coefficients $s_k$ ($k=0,1,2$) are determined by the eigenvalues $\lambda_k$ of $M_i$ (or equivalently of $M_i' = U^\dagger M_i U$). Explicitly one has
\begin{align}
s_0 &= -\frac{\lambda_1\lambda_2 e^{-\mathrm{i}\lambda_3 L_i/(2E_\nu)}}{(\lambda_2-\lambda_3)(\lambda_3-\lambda_1)}
        -\frac{\lambda_2\lambda_3 e^{-\mathrm{i}\lambda_1 L_i/(2E_\nu)}}{(\lambda_1-\lambda_2)(\lambda_3-\lambda_1)}
        -\frac{\lambda_1\lambda_3 e^{-\mathrm{i}\lambda_2 L_i/(2E_\nu)}}{(\lambda_1-\lambda_2)(\lambda_2-\lambda_3)}, \nonumber\\[6pt]
s_1 &= +\frac{(\lambda_1+\lambda_2) e^{-\mathrm{i}\lambda_3 L_i/(2E_\nu)}}{(\lambda_2-\lambda_3)(\lambda_3-\lambda_1)}
        +\frac{(\lambda_2+\lambda_3) e^{-\mathrm{i}\lambda_1 L_i/(2E_\nu)}}{(\lambda_1-\lambda_2)(\lambda_3-\lambda_1)}
        +\frac{(\lambda_1+\lambda_3) e^{-\mathrm{i}\lambda_2 L_i/(2E_\nu)}}{(\lambda_1-\lambda_2)(\lambda_2-\lambda_3)}, \nonumber\\[6pt]
s_2 &= -\frac{e^{-\mathrm{i}\lambda_3 L_i/(2E_\nu)}}{(\lambda_2-\lambda_3)(\lambda_3-\lambda_1)}
        -\frac{e^{-\mathrm{i}\lambda_1 L_i/(2E_\nu)}}{(\lambda_1-\lambda_2)(\lambda_3-\lambda_1)}
        -\frac{e^{-\mathrm{i}\lambda_2 L_i/(2E_\nu)}}{(\lambda_1-\lambda_2)(\lambda_2-\lambda_3)}.
\end{align}
The eigenvalues $\lambda_k$ ($k=1,2,3$) can be written in analytic form~\cite{Barger:1980tf,Zaglauer:1988gz,Xing:2000gg} as
\begin{align}
 \lambda_1 &=\frac{1}{3} x - \frac{1}{3}\sqrt{x^2-3 y}\left[z + \sqrt{3 \left( 1- z^2\right)}\right]\;,\nonumber\\ 
 \lambda_2 &=\frac{1}{3} x + \frac{1}{3}\sqrt{x^2-3 y}\left[z - \sqrt{3 \left( 1- z^2\right)}\right]\;, \nonumber\\
 \lambda_3 &=\frac{1}{3} x + \frac{2}{3} z\sqrt{x^2-3 y}\;.
\end{align}
where
\begin{align}
x &= \Delta m_{21}^{2} + \Delta m_{31}^{2} - A\,, \nonumber\\[6pt]
y &= \Delta m_{21}^{2} \Delta m_{31}^{2} - A \left[ \Delta m_{21}^{2} (1 - \sin^2\theta_{12} \cos^2\theta_{13}) + \Delta m_{31}^{2} (1 - \sin^{2}\theta_{13}) \right], \nonumber\\[6pt]
z &= \cos \left[ \frac{1}{3} \arccos \frac{2x^{3} - 9xy - 27A \Delta m_{21}^{2} \Delta m_{31}^{2} \cos^2\theta_{12} \cos^2\theta_{13}}{2(x^{2} - 3y)^{3/2}} \right]\;.
\end{align}

The full evolution operator for a baseline $L$ is obtained by multiplying the segment operators in the propagation order
\begin{equation}
 S^{\rm Vari}(L) \;=\; S_N S_{N-1} \cdots S_1 \,,
 \label{eq:S-vari}
\end{equation}
where the superscript ``Vari'' refers to the variable density profile. The corresponding survival probability is
\begin{equation}
 P_{ee}^{\rm Vari}(E_\nu,L) \;=\; \big|S^{\rm Vari}_{ee}(L)\big|^2 \,.
 \label{eq:Pee-vari-def}
\end{equation}
Note that an arbitrary matter-density profile $\rho(x)$ can be approximated with controllable accuracy by choosing a sufficiently large number of segments $N$, so that this semi-analytical treatment allows one to achieve the desired precision for realistic, position-dependent density profiles. This semi-analytical approach, in which the realistic crustal profile is represented by a finite number of constant-density segments, provides an accurate and efficient way to calculate the oscillation probabilities of neutrinos in a non-uniform medium. Note that this framework has been written for antineutrinos, and the corresponding expressions for neutrinos can be obtained by changing $U \to U^{*}$ and $A \to -A$ in the effective Hamiltonian of Eq.~(\ref{eq:Hamiltonian-matter}).

\section{Numerical results}
\label{sec:numerical}

In this section we present our numerical results for the matter effects on the
$\bar\nu_e \to \bar\nu_e$ survival probability along the TS--JUNO and YJ--JUNO
baselines. We first quantify the impact of variable matter density on the
oscillation probability, and then discuss its implications for the
precision measurement of oscillation parameters and for the determination of
the neutrino mass ordering.

To start, we shall employ the current global best-fit values of the three-flavor
oscillation parameters, based on the NuFIT~6.0 (2024)
analysis~\cite{Esteban:2024eli}.
For the JUNO baselines, we consider the locations of the JUNO site as well as the YJ and TS nuclear power plants~\cite{JUNO:2015zny}. The TS–JUNO and YJ–JUNO baselines are determined based on the average positions of the individual reactor cores. For the Taishan complex, we use $L_{\rm TS} = 52.695~\mathrm{km}$, which represents the average distance from the first and second cores to JUNO. For the Yangjiang complex, we adopt $L_{\rm YJ} = 52.465~\mathrm{km}$, corresponding to the average distance from the third and fourth cores to JUNO.

The matter density along each baseline is described by the segmented piecewise-constant profiles constructed in Sec. \ref{sec:matter-profile}, specifically utilizing the sets {$\{\rho_i^{(0)},L_i^{(0)}\}$} introduced in Sec.~\ref{subsec:geo-model} and Sec.~\ref{subsec:profile-fluct}, including the initial air segment where $\rho=0$.
The corresponding baseline-averaged densities are defined in Eq.~(\ref{eq:rho-avg-L-only}). Using the segment information presented in Tables 1 and 2, we find 
\begin{equation}
\rho^{\rm Average}_{\rm TS} \simeq 2.541~\mathrm{g/cm^3}\;,
\label{eq: rho_A_TS}
\end{equation}
and
\begin{equation}
\rho^{\rm Average}_{\rm YJ} \simeq 2.554~\mathrm{g/cm^3}\;,
\label{eq: rho_A_YJ}
\end{equation}
which will serve as benchmark constant densities for the subsequent analysis.
We also specified conservative fluctuation ranges of $\pm 10\%$ for both the segment densities and segment lengths as discussed in Sec.~\ref{subsec:profile-fluct}.

\begin{figure}[H]
    \centering
    \includegraphics[width=0.9\linewidth]{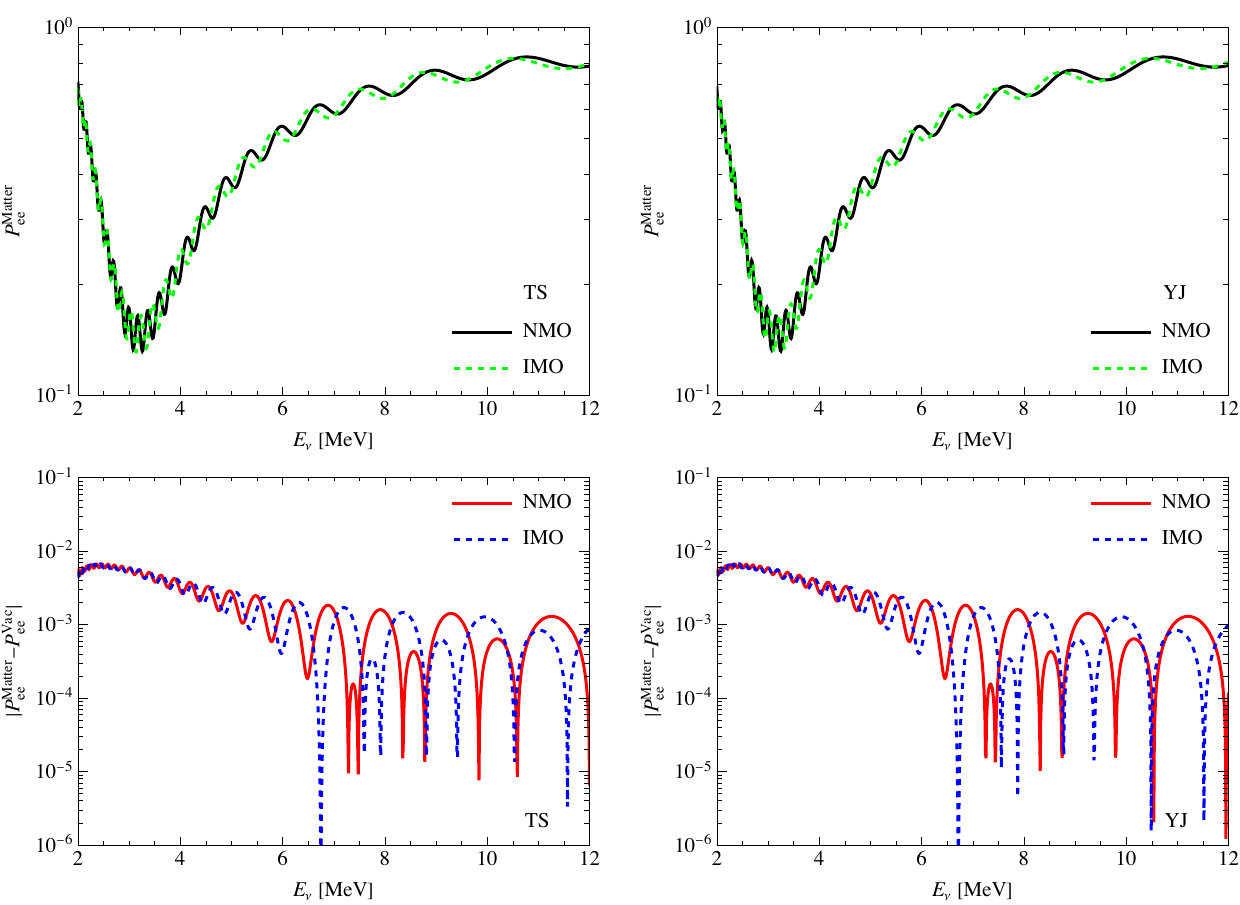}
    \caption{Top row: the electron antineutrino survival probability in matter, $P^{\rm Matter}_{ee}$,
as a function of $E_\nu$ for the TS--JUNO (left) and YJ--JUNO (right) baselines.
Bottom row: the absolute matter-induced correction with respect to vacuum, $|P^{\rm Matter}_{ee}-P^{\rm Vac}_{ee}|$.
Solid and dashed curves correspond to normal (NMO) and inverted (IMO) mass ordering, respectively.
The survival probabilities are calculated using the current best-fit values of the oscillation parameters~\cite{Esteban:2024eli}
and the piecewise-constant matter-density profiles obtained from the geological sampling along the baselines
(see Figs.~\ref{fig:TStoJUNO} and~\ref{fig:YJtoJUNO} and Tables~\ref{tab:segments-TS} and \ref{tab:segments-YJ}).
}
    \label{fig:VacMatterCom}
\end{figure}

First, we examine the absolute magnitude of variable matter effects on the
$\bar{\nu}_e \to \bar{\nu}_e$ survival probability. As shown in Fig.~\ref{fig:VacMatterCom}, the upper panels
display the variable-matter survival probability $P^{\rm Matter}_{ee}$ as a function of $E_\nu$
for the reference density profiles along the TS--JUNO (left) and YJ--JUNO (right) baselines,
while the lower panels show the absolute matter-induced correction with respect to vacuum,
$|P^{\rm Matter}_{ee}-P^{\rm Vac}_{ee}|$. One can see that $P^{\rm Matter}_{ee}$ itself is typically
$\mathcal{O}(10^{-1}\!-\!1)$ over the JUNO energy window, whereas the correction
$|P^{\rm Matter}_{ee}-P^{\rm Vac}_{ee}|$ stays at the subpercent level in absolute size, corresponding
to a relative spectral distortion at the $\mathcal{O}(1\%)$ level. Such a size is comparable to
the expected uncertainties in extracting $\Delta m^2_{21}$ and $\sin^2\theta_{12}$, and is also
consistent with Ref.~\cite{Li:2016txk} based on the constant-density approximation.


Next we will discuss the effects of variable density profiles. Considering the potential correlations in the density fluctuations of connected or nearby segments, we analyze two extreme scenarios:
\begin{itemize}
  \item Scenario A: profiles in which the densities of all rock segments are coherently rescaled within a permitted $\pm 10\%$, corresponding to global upward or downward shifts of the baseline-averaged density;
  \item Scenario B: profiles in which the densities of all rock segments are randomly rescaled within the allowed $\pm 10\%$ band, thus fine structures of the density variations are allowed.
\end{itemize}
All these sampled representative profiles satisfy $|\delta\rho_i|\leq 10\%$ and
$|\delta L_i|/L_i^{(0)}\leq 10\%$ for every segment, and their
baseline-averaged densities lie within a few-percent band around the
reference values, consistent with the quantitative estimates in
Sec.~\ref{subsec:profile-fluct}.

\begin{figure}[H]
    \centering
    \includegraphics[width=0.9\linewidth]{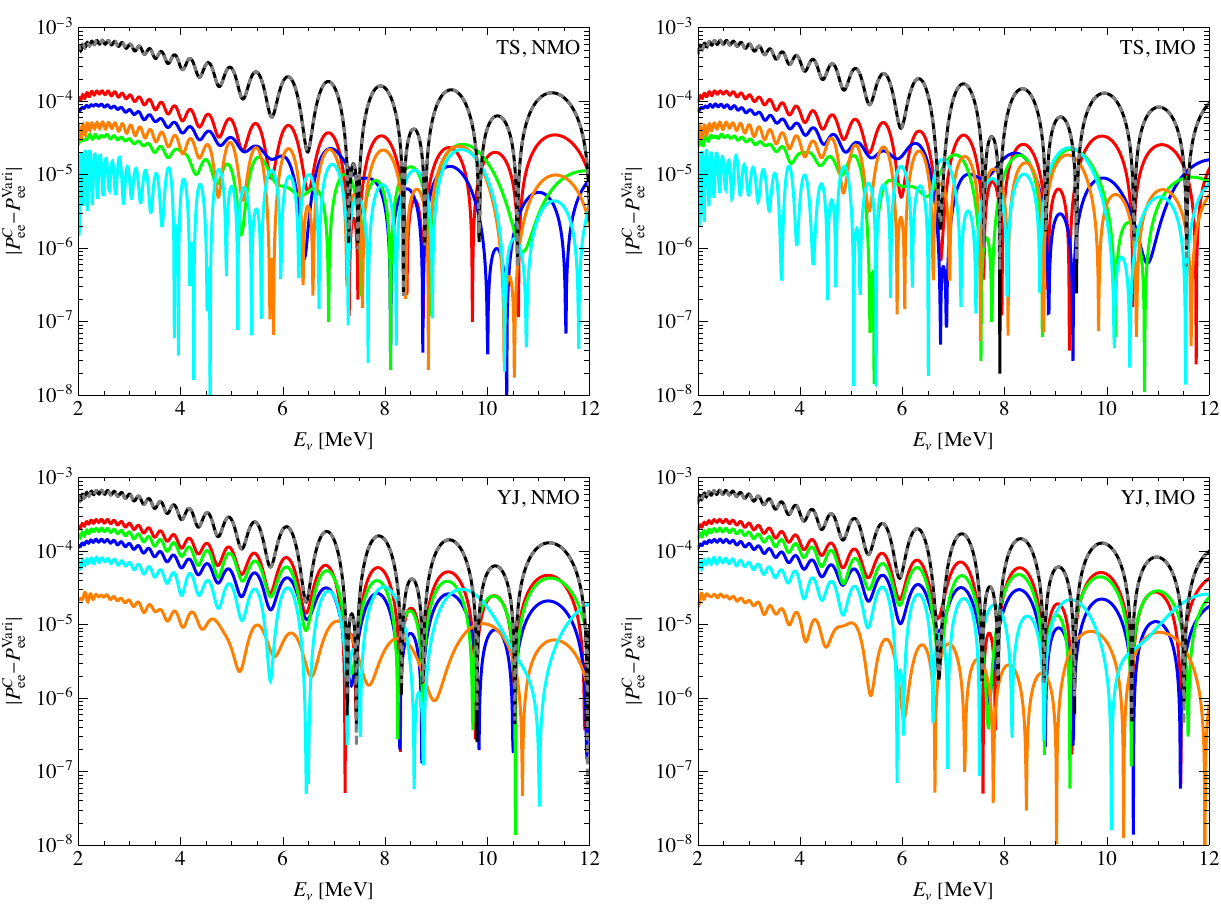}
\caption{Variable matter effects on the $\bar\nu_e \to \bar\nu_e$ survival probability along the TS--JUNO (upper row) and YJ--JUNO (lower row) baselines for normal (NMO, left column) and inverted (IMO, right column) mass ordering. Shown is the absolute difference between the probabilities computed with the benchmark constant-density profile, whose baseline-averaged densities $\rho^{\rm Average}_{\rm TS}$ and $\rho^{\rm Average}_{\rm YJ}$ are defined in Eqs.~\eqref{eq: rho_A_TS} and \eqref{eq: rho_A_YJ}, and with variable matter density, $|P^{\rm C}_{ee} - P^{\rm Vari}_{ee}|$. The black and gray curves correspond to Scenario~A, where all rock densities are coherently rescaled by $\pm 10\%$, while the colored curves show five representative realizations of Scenario~B with independent $\pm 10\%$ fluctuations in each rock segment.}
    \label{fig:Vari-proba}
\end{figure}

Using the benchmark constant density profiles in Eqs. \eqref{eq: rho_A_TS} and Eq. \eqref{eq: rho_A_YJ}, we compare the effects of variable matter against the corresponding constant approximation in Fig. \ref{fig:Vari-proba} for both the TS-JUNO and YJ-JUNO baselines. In each panel, the black and gray curves represent Scenario A, where the constant reference density is coherently increased or decreased within a $10\%$ range, respectively. These curves serve as a simple benchmark for the effect of fully correlated uncertainty in the reference crustal density profile, which is on the order of $10^{-3}$, in contrast to the total contributions at the percent level.
In contrast, the colored curves illustrate five representative realizations of fluctuations in Scenario B, generated by allowing the density in each rock segment along the baseline to vary independently within the $10\%$ range. It is important to note that the differences in survival probability are significantly smaller in Scenario B than in Scenario A due to the effects of statistical averaging.

Now we discuss how the matter-density variations affect the measurement of oscillation parameters and mass ordering.
From the first reactor antineutrino oscillation result of JUNO~\cite{JUNO:2025gmd}, the precisions of $\Delta m_{21}^2$ and $\sin^2\theta_{12}$ reached 1.55\% and 2.81\% respectively. With more data, these uncertainties are expected to be reduced to about 0.3\% and 0.5\% after 6 years of running, and to even smaller values over the full 20-year lifetime. 

\begin{figure}[H]
    \centering
    \includegraphics[width=0.9\linewidth]{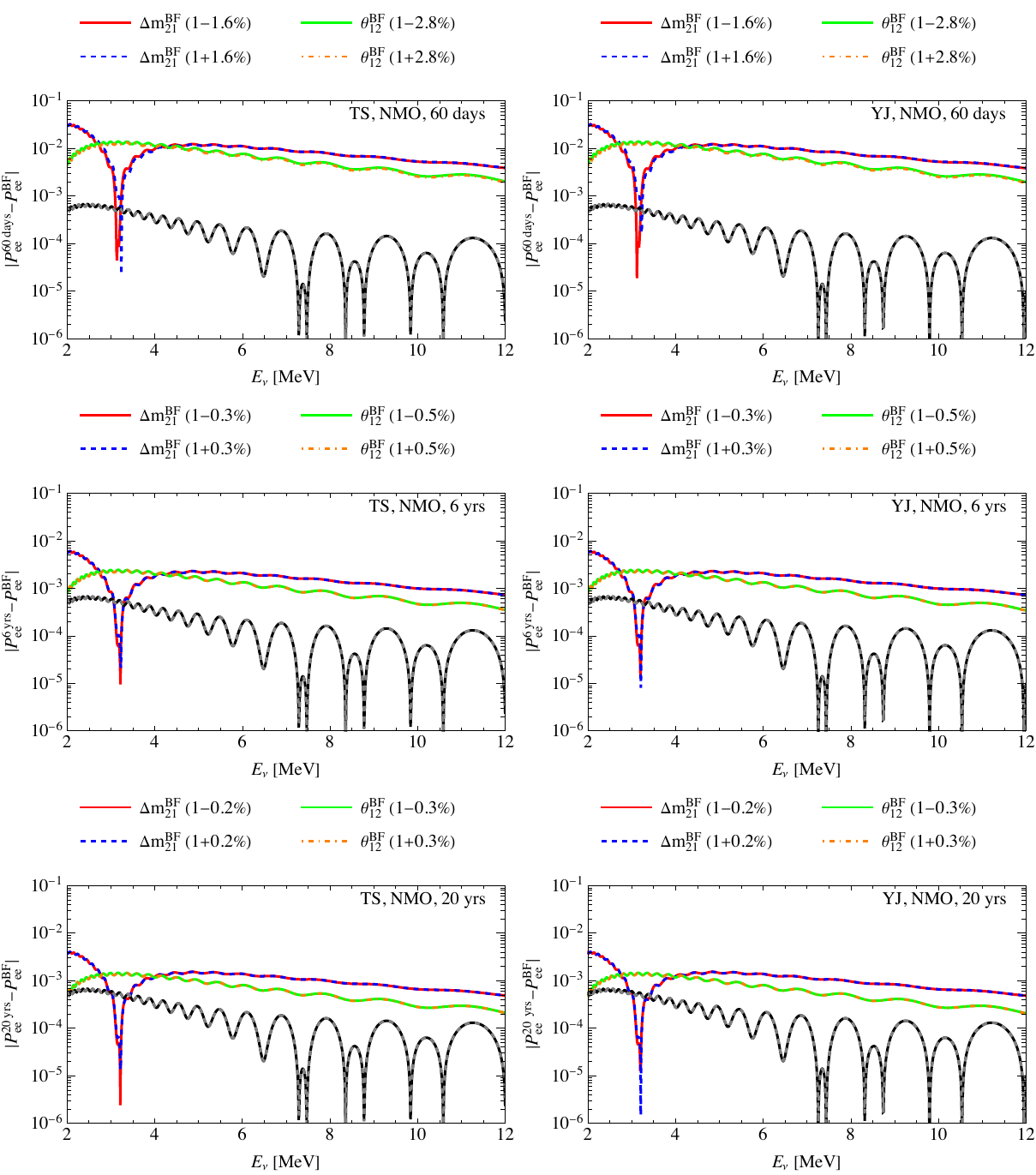}
\caption{Impact of the uncertainties in $\Delta m^2_{21}$ and $\theta_{12}$ on the $\bar\nu_e \to \bar\nu_e$ survival probability along the TS--JUNO (left column) and YJ--JUNO (right column) baselines for normal mass ordering. From top to bottom, the panels correspond to the expected parameter uncertainties after 60 days, 6 years (6 yrs), and 20 years (20 yrs) of data taking. In each panel, only one parameter is varied within its projected $1\sigma$ range, while the others are fixed at their best-fit values. The black and gray lines indicate the effect of a $\pm 10\%$ variation of the benchmark constant density, $\rho^{\rm Average}_{\rm TS}$ or $\rho^{\rm Average}_{\rm YJ}$ (Scenario~A), as in Fig.~\ref{fig:Vari-proba}.}
    \label{fig:NMOPara}
\end{figure}

In Fig.~\ref{fig:NMOPara}, the comparison of the survival probability differences between the oscillation parameter variation and the density profile variation is illustrated. As an illustration, Fig.~\ref{fig:NMOPara} shows the absolute differences in the survival probability, $\big|P_{ee}^{X} - P_{ee}^{\rm BF}\big|$
for the TS--JUNO baseline and NMO, where $P_{ee}^{\rm BF}$ is the probability calculated with the best-fit values, and $P_{ee}^{X}$ with one of the parameters shifted according to the expected JUNO precision. For clarity, in each panel only one parameter is varied at a time, while the others are fixed at their best-fit values.
The black and gray lines are that of variable density profile with the uncertainty of Scenario A. From the figures, one can read that the uncertainty of matter density profiles 
is currently more than one order of magnitude smaller than the uncertainty due to the oscillation parameters, but will grow to about $20\%$ (after 6 years) and $50\%$ (after 20 years) of the parameter-induced effect as the statistical precision improves.
 Therefore a careful investigation and modeling of the density profiles is needed in the future.

\begin{figure}[H]
    \centering
\includegraphics[width=0.9\linewidth]{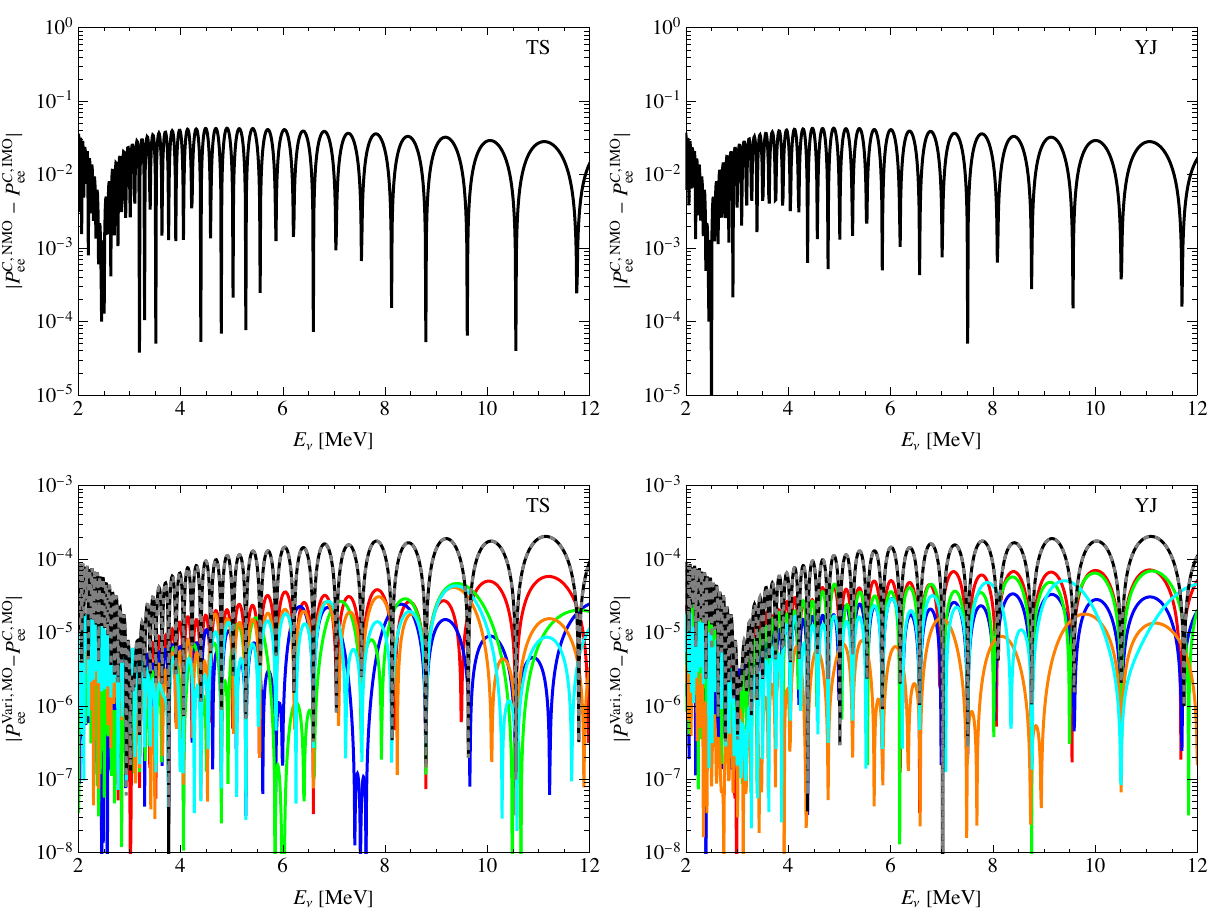}
    \caption{Variable matter effects on the determination of the neutrino mass ordering. Top row: absolute difference
$P^{\rm C,MO}_{ee} \equiv |P^{\rm C,NMO}_{ee} - P^{\rm C,IMO}_{ee}|$ between the constant-density survival probabilities in the normal mass ordering (NMO) and inverted mass ordering (IMO) for the TS--JUNO (left) and YJ--JUNO (right) baselines. Bottom row: deviation $|P^{\rm Vari,MO}_{ee} - P^{\rm C,MO}_{ee}|$ induced by matter-density variations. In all panels, the oscillation probabilities are computed using the current best-fit values of the oscillation parameters. The constant-density probabilities $P^{\rm C}_{ee}$ are obtained with the benchmark baseline-averaged densities $\rho^{\rm Average}_{\rm TS}$ and $\rho^{\rm Average}_{\rm YJ}$ defined in Eqs.~\eqref{eq: rho_A_TS} and \eqref{eq: rho_A_YJ}. The black and gray curves correspond to Scenario~A, while the colored curves represent five representative realizations of Scenario~B.}
    \label{fig:MO-Diff}
\end{figure}

We now turn to the impact of variable matter density profiles on mass ordering determination. In the upper panels of Fig. \ref{fig:MO-Diff}, we illustrate the differences in survival probabilities between NMO and IMO as follows:
\begin{equation}
 P_{ee}^{\rm C,MO}(E_\nu) \;=\; \big|P_{ee}^{\rm C,NMO}(E_\nu) - P_{ee}^{\rm C,IMO}(E_\nu)\big|\,,
\end{equation}
for the constant density profiles.
Similarly, we can define the differences in survival probabilities for variable density profiles as:
\begin{equation}
 P_{ee}^{\rm Vari,MO}(E_\nu) \;=\; \big|P_{ee}^{\rm Vari,NMO}(E_\nu) - P_{ee}^{\rm Vari,IMO}(E_\nu)\big|\,,
\end{equation}

\noindent and illustrate the absolute differences of these quantities in the lower panels of Fig. \ref{fig:MO-Diff}. Note that the black and gray curves represent Scenario A, while the colored curves correspond to Scenario B.
By comparing the magnitudes of these differences, we conclude that variations in matter density profiles contribute only $1\%$ to the mass ordering sensitivity, making them relatively negligible. Therefore, we can assert that the additional uncertainty introduced by the imperfect knowledge of the matter density profile is insignificant, and the determination of mass ordering remains robust against realistic variable matter density effects.

To put the matter-induced corrections discussed above into perspective, it is useful to compare them with
other uncertainties relevant for reactor antineutrino measurements.
For JUNO, detector- and reactor-related uncertainties can affect the predicted antineutrino rate at the
$\sim 1\%$ level~\cite{JUNO:2025gmd}.
On the other hand, the inverse beta-decay (IBD) cross section can be evaluated with sub-percent (typically per-mille) accuracy in state-of-the-art treatments~\cite{Ricciardi:2022}.
In comparison, the overall matter-induced modification of the survival spectrum shown in Fig.~\ref{fig:VacMatterCom} is at the percent level in absolute size,
while the additional impact from replacing the constant-density approximation by realistic, density profiles affects the mass-ordering discriminator only at the much smaller level.
Therefore, density-profile variations constitute a subleading effect with respect to the dominant experimental systematics, but they remain relevant for future analyses aiming at per-mille precision.

\section{Impact of anomalous density perturbations}
\label{subsec:anomalous-density}

In previous discussions, we have assumed that the matter density along the TS-JUNO and YJ-JUNO baselines can be described by a piecewise-constant profile with moderate segmented fluctuations, typically on the order of $\pm 10\%$ around the reference rock densities. This approach is well justified by the geological information available along the JUNO baselines.

In this section, however, we will explore the possibility of shallow crust tomography using reactor antineutrinos, considering the potential existence of localized structures that may deviate more significantly from the average crustal rock. These structures could include water-filled regions, underground reservoirs, or localized low-density cavities within fractured rocks.

For definiteness, let us consider a single localized anomaly with length $\Delta L_a$ and density
$\rho_a$ embedded in an otherwise standard rock profile characterized by an average density
$\bar{\rho}_0$.

If the baseline length is $L$, the new baseline-averaged density becomes

\begin{equation}
  \bar{\rho}' = \frac{1}{L}\left[ \rho_a \, \Delta L_a + \bar{\rho}_0 \, (L - \Delta L_a) \right]
  = \bar{\rho}_0 + \delta\bar{\rho} \,,
  \label{eq:rho-prime-anomaly}
\end{equation}
with the relative shift of
\begin{equation}
  \frac{\delta\bar{\rho}}{\bar{\rho}_0}
  = \left( \frac{\rho_a}{\bar{\rho}_0} - 1 \right) \frac{\Delta L_a}{L} \,.
  \label{eq:delta-rho-anomaly}
\end{equation}

This straightforward expression indicates that, even if $\rho_a$ differs significantly from the typical rock density, its impact on the baseline-averaged density is suppressed by the ratio $\Delta L_a/L$.
To quantify this effect, we take $\bar{\rho}_0 \simeq 2.6~\text{g/cm}^3$ as a representative crustal density and
$L \simeq 53~\text{km}$ as a typical baseline. A water-like anomaly corresponds roughly to
$\rho_a \simeq 1.0~\text{g/cm}^3$, while an air-filled cavity has $\rho_a \simeq 0$ on the scale relevant for
matter effects. 
For a $1~\text{km}$-long water anomaly one finds
\begin{equation}
  \frac{\delta\bar{\rho}}{\bar{\rho}_0}
  \simeq \left( \frac{1.0}{2.6} - 1 \right)
  \frac{1~\text{km}}{53~\text{km}}
  \approx -1.2\% \,,
  \label{eq:delta-rho-water}
\end{equation}
and for an air cavity of the same length,
\begin{equation}
  \frac{\delta\bar{\rho}}{\bar{\rho}_0}
  \simeq (0 - 1)
  \frac{1~\text{km}}{53~\text{km}}
  \approx -1.9\% \,.
  \label{eq:delta-rho-air}
\end{equation}
Shorter or more localized structures (e.g., \ $\Delta L_a \lesssim 0.5~\text{km}$) lead to relative shifts of
$\lesssim 1\%$, while more extended anomalous regions are disfavored by the available geological
information along the baselines.

The matter potential in the flavor basis is proportional to the electron number density,
\begin{equation}
  A = 2\sqrt{2}\, G_F N_e E_\nu \propto \rho \,,
  \label{eq:matter-potential-A}
\end{equation}
so that the fractional change of the effective potential is $\delta A / A \simeq \delta\bar{\rho}/\bar{\rho}_0$.
At the JUNO or future reactor antineutrino experiments, the standard matter effect induces only a percent-level modification of the
$\bar{\nu}_e$ survival probability with respect to vacuum oscillations. Therefore, an additional
$\mathcal{O}(1\%)$ change in the average density from a localized water-like or air-like anomaly leads
to a further suppression of the matter effect by a few $\times 10^{-4}$ in the survival probability.

\begin{figure}[t]
  \centering
\includegraphics[width=0.6\textwidth]{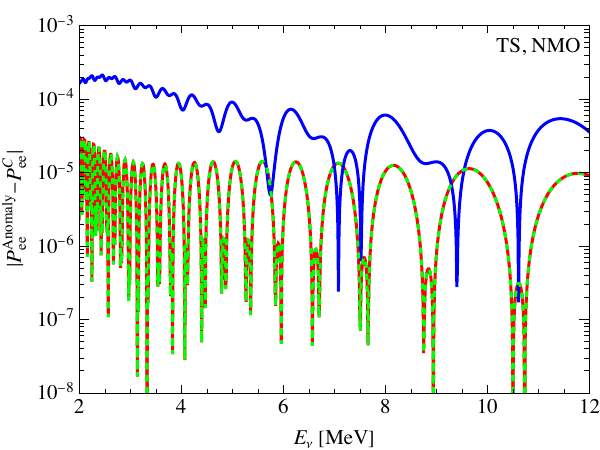}
  \caption{
    Absolute difference $|P_{ee}^{\rm Anomaly} - P_{ee}^{\rm C}|$ as a function of antineutrino
    energy $E_\nu$ for the TS--JUNO baseline in NMO. The constant-density
    reference probability $P_{ee}^{\rm C}$ is computed for a uniform matter density
    $\rho^{\rm Average}_{\rm TS}$, equal to the baseline-averaged density defined in Eq.~\eqref{eq: rho_A_TS}. The three
    colored curves correspond to profiles in which an additional $1~\text{km}$ air segment
    ($\rho \simeq 0$) is inserted at the beginning (red), in the middle (blue), or at the end (green)
    of the baseline, while the remaining path is kept at density $\rho^{\rm Average}_{\rm TS}$. The
    differences stay at the level of a few $\times 10^{-4}$ or smaller over the entire JUNO energy
    range, illustrating the negligible impact of such localized low-density anomalies on the
    reactor antineutrino survival probability.
  }
  \label{fig:TS-anomaly}
\end{figure}

To make this conclusion more explicit, Fig.~\ref{fig:TS-anomaly} illustrates, for the TS--JUNO baseline
and NMO, the absolute difference of
$|P_{ee}^{\rm Anomaly} - P_{ee}^{\rm C}|$ obtained in a constant-density
profile and in profiles containing a localized air-like segment. The constant reference probability
$P_{ee}^{\rm C}$ is computed assuming a uniform density $\rho^{\rm Average}_{\rm TS}$, and in the anomalous cases we keep the same uniform
density $\rho^{\rm Average}_{\rm TS}$ along the whole baseline, except for an additional
$1~\text{km}$ air layer (with $\rho \simeq 0$) inserted either at the beginning of the baseline
(red curve, between $0$ and $1~\text{km}$), in the middle (blue curve, around $L/2$), or close to
the detector (green curve, near the end of the baseline). Over the full neutrino energy window
($E_\nu \simeq 2$--$12~\text{MeV}$), the induced differences remain at the level of a few
$\times 10^{-4}$, in excellent agreement with the analytical estimate based on
Eqs.~(\ref{eq:delta-rho-anomaly})--(\ref{eq:delta-rho-air}).
We have also verified that analogous configurations for the YJ--JUNO baseline and for IMO lead to very similar corrections. In addition replacing the anomalous air layer by a water-like segment with $\rho_a \simeq 1.0~\text{g/cm}^3$ yields very similar results.

More generally, we have explicitly examined several extreme configurations in which a rock segment in the baseline model is replaced by either water or air, with $\Delta L_a$ extending up to the kilometer scale and located at various positions along the baseline. In all these cases, the resulting changes in $P(\bar{\nu}_e \to \bar{\nu}_e)$ were found to be much smaller than the effects of the typical $\pm 10\%$ segment-wise fluctuations discussed in Sec. 4. Therefore, we conclude that potential anomalous density structures, such as localized water-filled regions or air cavities, do not undermine the robustness of our conclusions regarding matter effects on reactor antineutrino oscillations at JUNO. As a result, an even larger detector is necessary for effective shallow crust tomography using reactor antineutrinos.

\section{Summary}
\setcounter{equation}{0}

\noindent
In this work, we have investigated the impact of realistic matter density variations on reactor antineutrino oscillations. Using geological information along the baselines from the Taishan and Yangjiang nuclear power plants to the experimental hall, we constructed a piecewise-constant density model that approximates the South China crustal structure. We then allowed for conservative fluctuations of $\pm10\%$ in the density of each segment and $\pm10\%$ in the segment lengths, while keeping the total baseline lengths fixed. The resulting average matter densities and the conservative uncertainties are given as
\begin{equation*}
\rho^{\rm Average}_{\rm TS} \simeq (2.541\pm10\%)~\mathrm{g/cm^3}\;,
\end{equation*}
and
\begin{equation*}
\rho^{\rm Average}_{\rm YJ} \simeq (2.554\pm10\%)~\mathrm{g/cm^3}\;.
\end{equation*}

Within the standard three-flavor oscillation framework, we computed the $\bar\nu_e \to \bar\nu_e$ survival probabilities for both constant-density and variable-density profiles, and quantified the resulting differences for both normal and inverted mass orderings. We further assessed the implications of these differences on the precision measurement of the oscillation parameters $\Delta m_{21}^2$ and $\sin^2\theta_{12}$ and neutrino mass ordering.

Our main conclusions can be summarized as follows:

\begin{itemize}
 \item The absolute difference $|P_{ee}^{\rm C} - P_{ee}^{\rm Vari}|$ induced by realistic density variations is approximately $10^{-3}$ for the correlated density fluctuations of Scenario A, accounting for about 10\% of the total matter effects. While this effect is negligible at the current precision, it should be considered and addressed for future measurements aiming at per-mille $(10^{-3})$ precision.
 
 \item The mass-ordering discriminant $P_{ee}^{\rm C,MO} = |P_{ee}^{\rm C,NMO} - P_{ee}^{\rm C,IMO}|$ is typically on the order of $10^{-3}$--$10^{-2}$, whereas the correction $|P_{ee}^{\rm Vari,MO} - P_{ee}^{\rm C,MO}|$ resulting from variable matter density is at most on the order of $10^{-4}$. Consequently, realistic density variations may statistically alter the ordering discriminant by at most a few percent, which is negligible. Thus, the mass-ordering determination at JUNO is largely insensitive to the detailed structure of the crustal density along the baselines.
\end{itemize}
From a broader perspective, the commencement of JUNO's data collection and the release of the first reactor antineutrino oscillation results signify an important advancement in the global neutrino program. Alongside Hyper-Kamiokande and DUNE, JUNO will contribute to completing the understanding of three-flavor oscillations by determining the mass ordering and significantly enhancing the precision of the mixing angles and mass-squared differences.
With the advent of the era of precision measurements in neutrino physics, the contributions of sub-leading and subtle effects may need to be reconsidered, such as the variable matter density profile considered in this work. Other such efforts include studies of indirect unitarity violation effects~\cite{Li:2018jgd}, radiative corrections to neutrino propagation in matter and to elastic neutrino-electron scattering~\cite{Huang:2023nqf,Huang:2024rfb,Huang:2025apv,Huang:2025iww}. Systematically incorporating these effects will further increase the reliability of oscillation experiments, enhance their discovery potential~\cite{Schwetz:2021cuj, Wang:2025ess}, and support precise measurements of neutrino oscillation parameters and searches for related new physics.

\vspace{0.3cm}
\section*{Acknowledgments}
Jing-yu Zhu thanks Wen-jie Wu for helpful discussions.
The work of Jing-yu Zhu was supported in part by the National Key Research and Development
Program of China (under grant number 2021YFA1601300) and by the National Natural Science
Foundation of China (under grant number 12505134).
The work of Yu-Feng Li was supported in part by the National Natural Science Foundation of
China (under grant number 12075255).
The work of Andong Wang was supported in part by the Open Project Program of National Key Laboratory of Uranium Resources Exploration-Mining and Nuclear Remote Sensing (under grant
number NKLUR-2024-QN-005).
The work of Ya Xu was supported in part by the National Natural Science Foundation of China (under grant number 42074092).

\bibliographystyle{JHEP_improved.bst}
\bibliography{JUNO-varyM.bib}

\end{document}